# Dendritic magnetic instability in superconducting MgB$_2$ films

T. H. Johansen*, M. Baziljevich*, D. V. Shantsev*, P. E. Goa*, Y. M. Galperin*, W. N. Kang[§], H. J. Kim[§], E. M. Choi[§], M.-S. Kim[§], S. I. Lee[§]

\* *Department of Physics, University of Oslo, POB 1048 Blindern, N-0316 Oslo, Norway.*
§ *National Creative Research Initiative Center for Superconductivity, Department of Physics, Pohang University of Science and Technology, Pohang 790-784, Republic of Korea.*

**Following the discovery of superconductivity in polycrystalline magnesium diboride a tremendous effort is now focused on thin films of this material. Contrary to expectations, the penetration of magnetic flux in such films is found dominated by large dendritic structures abruptly created when small fields are applied. The dendritic instability, observed below 10 K using magneto-optical imaging, has a temperature dependent morphology ranging from quasi-1D dendrites to beautiful tree-like structures. This behavior is responsible for the anomalous noise in magnetization curves, and strongly suppresses the apparent critical current. Simulations of vortex dynamics incorporating local heating effect reproduce the observed dendritic scenario.**

The new superconductor, MgB$_2$, discovered *(1)* in January this year has already proved to be a promising candidate for technological applications due to success in fabrication of thin films *(2)* and wires *(3)* with high current carrying capabilities. At the same time, such films and wires, as well as polycrystalline MgB$_2$ are reported to show exceptional magnetic behavior displaying numerous and "noise-like" jumps in the magnetization as a function of applied field *(3-5)*. Magnetization jumps in type-II superconductors are usually associated with a thermo-magnetic instability of the quantized flux lines (vortices) penetrating the material. When the vortices move they leave a trail of elevated temperature facilitating motion of nearby vortices, which eventually leads to a large-scale avalanche invasion of depinned flux lines *(6)*. This thermal runaway, where the magnetic energy stored in the superconductor suddenly converts to thermal energy, can cause instability of the superconducting properties and have catastrophic consequences for practical applications. To which extent the thermo-magnetic instability will affect the technological potential of MgB$_2$ is today a vitally important question. In high temperature superconductors (HTSs) the flux jumps occur only in bulk materials, the first one at applied fields of typically one Tesla and then nearly periodically as the field increases. In MgB$_2$ the jumps are omnipresent, with the first one occurring already at a few milliTesla and subsequent jumps coming at random. Thus, MgB$_2$ appears not only far more susceptible to thermal runaways than HTSs, but the flux jumps exhibit also qualitatively new features. All this motivated the present study of MgB$_2$ films using magneto-optical (MO) imaging to visualize and characterize the nature of the magnetic instability.

Thin films of MgB$_2$ were fabricated on (1$\bar{1}$02) Al$_2$O$_3$ substrates using pulsed laser deposition. An amorphous B film was first deposited, and then sintered at high temperature in a Mg atmosphere. Details of the preparation are reported elsewhere *(2)*. Typical films had a sharp superconducting transition ($\Delta T_c \sim 0.7$ K) at $T_c = 39$ K, and a high degree of *c*-axis alignment perpendicular to the film plane.

Magnetic characterization of the films was first done by measuring the magnetization versus applied field (Fig. 1). The vertical width of such hysteresis loops represents the integral magnetic moment of the shielding currents, and is to a good approximation proportional to the critical current density, $J_c$. The magnetization was measured at various temperatures from 25 K and below, and as expected, the loop becomes wider the lower the temperature. Furthermore, peaks are present around $H = 0$ implying a substantial enhancement of $J_c$ at low fields. At 10 K the behavior evidently changes character as distinct "noisy" features start to appear in the central part of the magnetization curves. The fluctuations are also accompanied by a cutoff of the central peak, thereby decimating the favorable low-field value of $J_c$. The amplitude of the fluctuations is found to be largest near 10 K, while the central peak cutoff is more severe at lower temperatures. To reveal the details of this most unexpected behavior, found to be typical for such films, we performed space-resolved magnetic measurements. A sample of thickness 400 nm and size 5×5 mm$^2$ was chosen for MO-imaging studies.

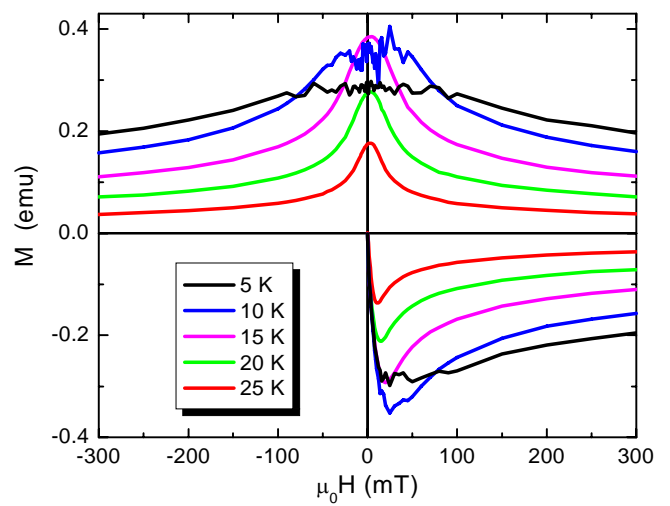

**Fig. 1.** Magnetization hysteresis loops for a MgB$_2$ thin film. The measurements were done at different temperatures between 25 and 5 K using a SQUID magnetometer (Quantum Design MPMS). Pronounced fluctuations in the magnetization are evident at $T = 10$ and 5 K, while only regular behavior is seen at the higher temperatures. The fluctuations indicate that numerous flux jumps are taking place. Their magnitude as well as the accompanying effect of cutting off (flattening) the central peak vary with the temperature.

Shown in Fig. 2 is a sequence of MO images taken during a field cycle at 5 K. Immediately one sees that the magnetic behavior of the MgB$_2$ film differs totally from the usual critical-state type of smooth flux penetration patterns. The images for increasing field (A)-(D), show that the flux instead penetrates in dendritic structures which, one after the other, invade the entire film area. The dendrites nucleate at seemingly random places near the film edge, and grow to their final size in less than 1 ms (the time resolution of our CCD system). Moreover, it is found that once a dendrite is formed its size remains constant although the applied field continues to increase. When the field is subsequently reduced (E,F), the flux redistributes in the same abrupt manner leading to a remanent state with an overlapping mixture of two types of dendrites – one containing trapped flux of initial polarity and one with antiflux due to penetration of the reverse return field of the trapped vortices. From these images, which may be directly compared to the black curve in Fig. 1, it is clear that the irregular features in the magnetization data stem from the formation of flux dendrites.

A striking change in the morphology of the flux patterns was observed when MO-imaging was carried out at different temperatures (Fig. 3 A-C). At the lowest temperature of 3.3 K the number of dendrites is large, and each dendrite has only a few branches. With increasing temperature the degree of branching grows, as seen in the image taken at 9.9 K, where a large tree-like structure was formed in one single burst of flux motion. These results reveal why the fluctuations in the magnetization data changed from numerous small jumps at low temperatures to fewer but larger jumps at 10 K. MO-imaging showed also that above 10 K the dendrite formation never occurs and is replaced by a regular behavior, also this fully consistent with the magnetization data.

This remarkable complexity in the flux dynamics suggests that an instability is taking place in MgB$_2$ films. The fact that unstable conditions exist only at the lowest temperatures points towards one of thermo-magnetic origin. It is well known that the heat released by vortices in motion causes a local temperature rise (large at low $T$ since the specific heat is small), which facilitates further flux motion that in turn can trigger an avalanche-like invasion. We have used this mechanism as basis for vortex dynamics simulations, where aspects particular for films in a perpendicular field are taken into account.

First, in superconducting films where the thickness is of the order of the London penetration depth, or smaller, the extreme demagnetization results in a Meissner sheet current that flows in the entire sample area. For a thin rectangular film the Meissner current flows *(7)* as shown by the stream lines in Fig. 4A, and the Lorentz force $F_M$ from this current on a vortex is directed as indicated by the arrows. From the overall alignment of the dendrites seen in the MO image (B), it is clear that the Meissner current is a major driving force in their formation. A second aspect included in the simulations is that in thin films the vortex-vortex repulsion $F_{ij}$ is long range *(8)*, and we use here the asymptotic $r_{ij}^{-2}$ dependence.

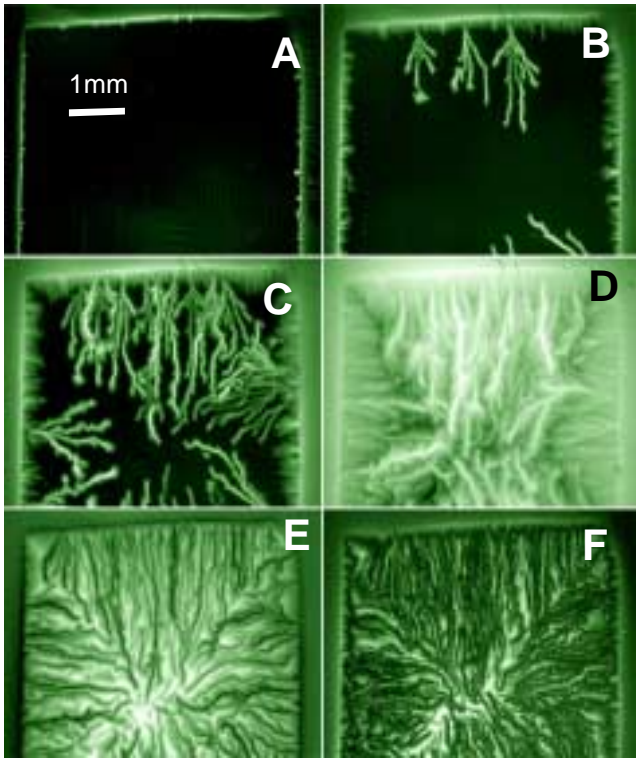

**Fig. 2.** MO images of flux penetration into the virgin state at 5 K (image brightness represents the flux density). **(A-D)** Images taken at applied fields (perpendicular to the film) of 3.4, 8.5, 17, 60 mT, respectively. **(E,F)** Images taken at 21 and 0 mT during the subsequent field reduction. Except for an initial stage, the flux penetration is strongly dominated by the abrupt appearance of dendritic structures nucleating at seemingly random places at the sample edge. The behavior is essentially independent of the field sweep rate.

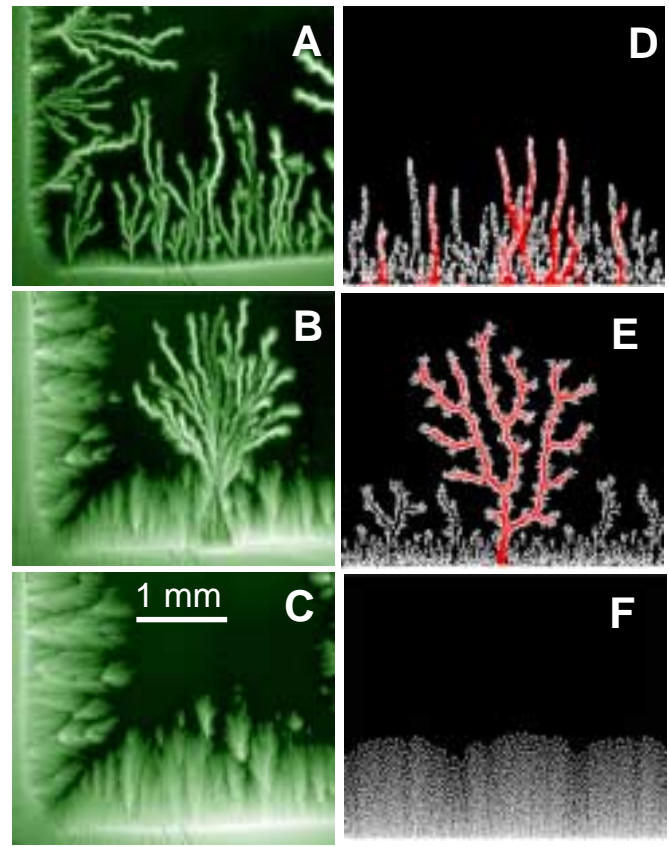

**Fig. 3.** Different types of flux pattern morphology at various temperatures. **(A-C)** MO images taken for $T$ = 3.3, 9.9 and 10.5 K at applied fields of 13, 17 and 19 mT, respectively. At low $T$ the dendrites are numerous and with few branches, while just below 10 K only large tree-like structures are formed. Above 10 K the film behaves traditionally according to the critical-state model. **(D-F)** Results of computer simulations largely reproducing the observed types of flux penetration patterns. Individual vortices are indicated by white dots, and red indicate elevated temperature due heat dissipated by the most recent vortex motion.

The pinning of the vortices was accounted for by assuming a position-independent pinning energy $U_{pin}$. Uniform pinning was motivated by the observation that when flux penetration experiments are repeated the detailed dendrite pattern is never recurring, i.e., the exact shape and number of dendrites vary at random, thus showing that this flux dynamics is insensitive to local variations in the pinning. Finally, it is assumed that a trail of elevated temperature follows the path of every moving vortex. More details of the simulations are given in *(9)*.

Seen in Fig. 3 are the results of simulations corresponding to three different temperatures. All the types of penetration pattern observed in the experiments are here reproduced; quasi 1-dimensional dendrites at low $T$ (D), one large and highly branching structure at intermediate $T$ (E), and the conventional smooth flux profile at high $T$ (F). Also the rapid dynamics is there, since avalanches are triggered by a small incremental step in the applied field, and followed by a fast growth of the full dendrite. Shown red in (D) and (E) are the regions where the recent bursts of flux motion have increased the temperature. In general, the most heated part is the core of a dendrite, reflecting large traffic of vortices. Due to high temperature the vortices move hastily through this region, and finally the core can end up empty of flux, see Fig. 5. Remarkably, this prediction could be confirmed experimentally, as our MO imaging was able to resolve several dendrite cores of low flux density.

Figure 6 summarizes our results on flux penetration in $MgB_2$ films. Dendritic flux structures appear only as the field exceeds a certain temperature dependent threshold value. The morphology of the dendrites is also changing with temperature. Similar behavior has been observed earlier in niobium films *(10-12)*, and in $YBa_2Cu_3O_7$ when triggered by a laser pulse *(13,14)*. Its understanding, however, was hindered because the existing theories for a thermo-magnetic instability *(6)* were developed only for bulk superconductors, and considered only a coarse, non-dendritic scenario. In thin films the scenario is totally different as the flux pattern is strongly branched and the threshold field much lower *(15)*. The thermo-magnetic origin of the dendritic instability is supported also by simulations based on the Maxwell and heat diffusion equations *(17)*.

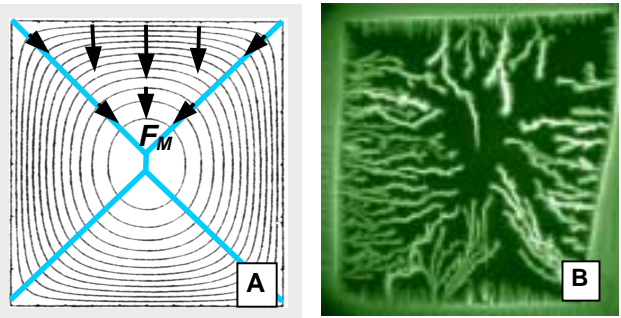

**Fig. 4.** **(A)** Current stream lines in a rectangular thin film superconductor in the Meissner state *(7)*. The arrows show the Lorentz force $F_M$ acting on a vortex. **(B)** MO image of the whole $MgB_2$ film. The global penetration of dendrites is governed by the Meissner current because (i) dendrites nucleate preferably where $F_M$ is maximum, (ii) dendrites grow along the direction of $F_M$, and (iii) they terminate where $F_M$ is small, i.e., near the center or along the diagonals. The MO image is one frame of a VIDEO sequence *(20)* recorded at 3 K as the field was ramped from 0 to 35 mT.

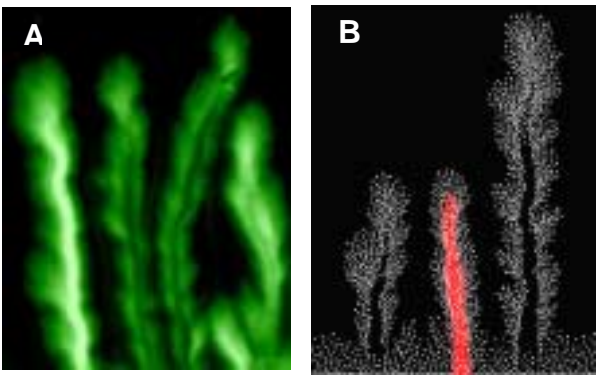

**Fig. 5.** Internal structure of dendrite branches. **(A)** MO image of dendrites where two of them have cores with low flux density. **(B)** Vortex dynamics simulations reproducing the experimental observation. Shown red is the heated core of a growing dendrite.

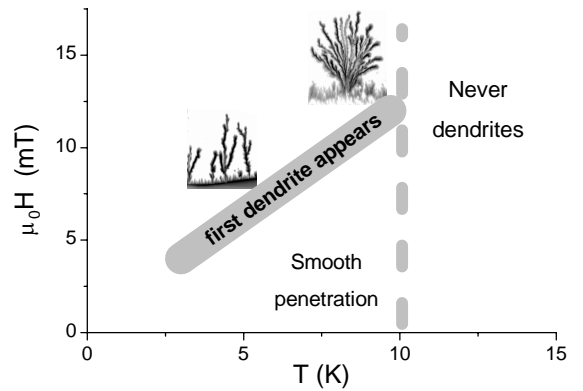

**Fig. 6.** Field-temperature diagram for onset of the flux dendrite instability in the present $MgB_2$ film. Other films with different thickness showed similar behavior although the threshold field and temperature varied.

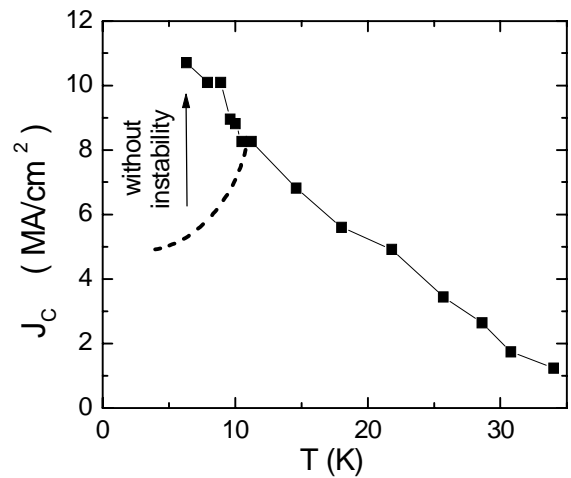

**Fig. 7.** "True" $J_c$ derived from MO images (squares) and apparent $J_c$ estimated from *M(H)* loops (dashed line) like those shown in Figure 1. From the MO images $J_c$ was obtained by the position of the flux penetration front at applied fields below the instability threshold. The Bean-model result *(7)* for a thin strip in a perpendicular field was used.

A key question for technological applications of $MgB_2$ is how common the flux jump instability is in this material. From reports published so far the answer may seem pessimistic. Dendritic flux penetration or magnetization jumps have been observed in a large number of $MgB_2$ samples: In the present high-quality films with $J_c \sim 10^7$ A/cm$^2$, and thickness $d$ = 100-400 nm, see also *(4)*, in polycrystalline samples *(5)* with $J_c \sim 10^5$ A/cm$^2$, $d \sim$ 1 mm, and in iron-clad wire *(3)* with $J_c \sim 10^5$ A/cm$^2$, $d \sim$ 35 µm. This wide range of $J_c$-values and sample types makes it clear that the flux jump instability is a strong habit of $MgB_2$ *(18)*. However, the problem can be fought using methods already developed for conventional (low-$T_c$) superconductors, i.e., by introducing thermal stabilization as done, e.g., for wires clad with normal metal.

The benefits of avoiding the instability in $MgB_2$ films can be illustrated by Fig. 7, showing $J_c$ from 35 K and down to temperatures where dendrites dominate the magnetic behavior. This curve represents the "true" low-field value of $J_c$, since the data were obtained from flux density profiles in MO images taken before any flux jump was seen. In today's films most global characteristics, such as magnetization, *I-V* curve and AC losses are actually governed by a much lower $J_c$, here indicated by the dashed line. A challenge for the future is to find practical ways to stabilize the magnetic breakdown, and re-establish the very high "true" critical current density in $MgB_2$ films.

**Acknowledgements**
We are grateful to A. Polyanskii, D. Larbalestier, E. Altshuler, A. Gurevich, P. Leiderer, U. Bolz and B.-U. Runge for fruitful discussions. The work was financially supported by The Norwegian Research Council, and by the Ministry of Science and Technology of Korea through the Creative Research Initiative Program.

Correspondence should be addressed to T. H. Johansen (e-mail: t.h.johansen@fys.uio.no).
Request of the materials should be addressed to S. I. Lee (e-mail: silee@postech.ac.kr